\documentclass[twocolumn,showpacs,preprintnumbers,amsmath,amssymb]
{revtex4}
\usepackage{subfigure}
\usepackage{graphicx,dcolumn,bm,color,latexsym,amssymb}
\usepackage[latin1]{inputenc}
\usepackage{import}
\usepackage{xifthen}
\usepackage{pdfpages}
\usepackage{transparent}
\usepackage{verbatim}

\definecolor{Blue}{rgb}{0.0,0.0,1}
\definecolor{Red}{rgb}{1,0.0,0.0}
\definecolor{Green}{rgb}{0,0.5,0.0}

\begin{document}

\title{Heat Distribution of Relativistic Brownian Motion} 

\author{Pedro V. Paraguass\'{u}}
 \email{paraguassu@aluno.puc-rio.br}
\affiliation{Departamento de F\'{i}sica, Pontif\'{i}cia Universidade Cat\'{o}lica\\ 22452-970, Rio de Janeiro, Brazil}

\author{Welles A.~M. Morgado}
 \email{welles@puc-rio.br}
\affiliation{Departamento de F\'{i}sica, Pontif\'{i}cia Universidade Cat\'{o}lica\\ 22452-970, Rio de Janeiro, Brazil\\ and National  Institute of Science and Technology for Complex Systems}

\date{\today}

\begin{abstract}
Understanding the statistical behavior of the heat in stochastic systems gives us insight about the thermodynamics of such systems. Using the recently proposed Relativistic Stochastic Thermodynamics, we investigate the statistics of the heat of a Relativistic Ornstein-Uhlenbeck particle, comparing with the classical cases. The results are exact through numerical integration of the Fokker-Planck of the joint distribution, and are validated by numerical simulations.  
\end{abstract}


\maketitle



%
%
\section{Introduction}

Understanding the behavior of the heat exchanged by a system has always had an important role in physics. Since the beginnings of thermodynamics, understanding how a system loses or gains energy from the surroundings was important to develop early thermal machines \cite{carnot2012reflections}. With today's technological advances, we are able to thermodynamically interact on an increasingly smaller scale, ranging from micrometer to nanometer. Typically, these systems are far from equilibrium, where the thermodynamics's functionals, such as heat, entropy, or work, are treated as fluctuating quantities. The field of the investigations of the thermodynamics of such fluctuating systems is well known in the literature as the Stochastic Thermodynamics~\cite{oliveira2020classical,ciliberto_experiments_2017,seifert2012stochastic,sekimoto2010stochastic}.

Heat is a fundamental quantity in Stochastic Thermodynamics, i.e.,  the energy naturally exchanged between the system and the surrounding, in a disordered way. As a random variable, characterization of the statistics of heat for diffusive systems was carried in many different models \cite{paraguassu_heat_2021_2,paraguassu_heat_2021,gupta_heat_2021,fogedby_heat_2020,goswami_heat_2019,crisanti_heat_2017,ghosal_distribution_2016,rosinberg_heat_2016,kim_heat_2014,kusmierz_heat_2014,saha_work_2014,chatterjee_single-molecule_2011,chatterjee_exact_2010,fogedby_heat_2009,imparato_probability_2008,imparato_work_2007,joubaud_fluctuation_2007}.  These works bring physical insights into the thermodynamics of classical diffusive systems, however, as far as we known, they only deals with non-relativistic systems.  

Relativistic diffusive systems are expected to be found in nature, and can appear in quite different phenomena such as cosmic jets \cite{pal_stochastic_2020, meyer2018cosmic}, quark-muon plasma produced by heavy ion collisions \cite{koide_thermodynamic_2011,akamatsu_heavy_2009}, and graphene in a semiclassical regime ~\cite{pototsky_relativistic_2012,pototsky_periodically_2013}. From a more mathematical point of view, these relativistic diffusive systems can be  modeled by the Relativistic Brownian motion~\cite{dunkel_theory_2005,dunkel_theory_2005-1,dunkel_relativistic_2009} which are a particular case of Brownian motion with nonlinear friction term \cite{lindner_diffusion_2007}. One well studied case is the Relativistic Orstein-Uhlenbeck \cite{debbasch_relativistic_1997}, where its relaxation properties were studied in \cite{debbasch_thermal_2012,felderhof_momentum_2012}.

Recently, Pal and Deffner proposed a Stochastic Thermodynamic framework for the Relativistic Brownian motion \cite{pal_stochastic_2020}. Their model is based on the Relativistic Ornstein-Uhlenbeck case, where they define heat, work, and entropy for the relativistic Brownian particle. This relativistic version of classical Stochastic thermodynamics quantities~\cite{sekimoto2010stochastic} was shown to satisfy the conservation of energy and the fluctuation theorem version of the second law of thermodynamics. Since heat is a fundamental quantity, we thus want to understand its behavior for this new relativistic case. It is meaningful to check if the relativistic stochastic thermodynamics leads to a consistent statistical behavior of the heat. Moreover, it is also interesting to compare with the classical case. How different is the behavior of the heat between the classical and relativistic system?  In addition, it is important to notice, that the same definition of heat was also proposed by Koide and Kodama in~\cite{koide_thermodynamic_2011}.

In the present paper, by means of the Stochastic Thermodynamics framework \cite{pal_stochastic_2020}, we investigate the heat distribution for the Relativistic Ornstein-Uhlenbeck model. We obtain the heat distribution in two distinct limits, the exact relativity limit, and the ultra-relativistic limit. The results are exact through numerical integration. We use the variational formula for the Fokker-Planck \cite{langtangen2016solving,risken1996fokker}, and we integrate it numerically. We also use the Path integral formalism \cite{onsager1953fluctuations,moreno_conditional_2019,wio2013path,chaichian2018path} to deal with the ultra-relativistic case. The results are compared with numerical simulations of the stochastic process and are found in agreement. 

The paper is organized as follows: In section \ref{sec2} we define the model and its thermodynamics. In section \ref{sec3} we studied the heat fluctuations of the relativistic case by path integrals. In section \ref{sec4} we obtain, through numerical integration of the Fokker-Planck equation of the joint distribution. We conclude in section \ref{sec5} with a discussion of the results.

\section{Relativistic Orstein-Uhlenbeck} \label{sec2}

As a first model to study the heat fluctuations of a relativistic particle, we study the free case, often called Relativistic Ornstein-Uhlenbeck \cite{dunkel_relativistic_2009}. Despite being the most simple situation for the relativistic Brownian particle, we will see that the non-linear dependency on the momentum can lead us to non-trivial results. 

A relativistic particle is a particle with an  absolute velocity not exceeding the speed of light $|V(t)|<c$.  Its stochastic behavior  for the momentum is defined in the inertial frame of the environment and the corresponding stochastic equation is (unless explicitly stated, we assume $c=1$)
\begin{equation}
    dp = - \Gamma(p)dt +\sqrt{2\gamma MT}dW_t, \label{langevin}
\end{equation}
where \[\Gamma(p)= \gamma M \frac{p}{\sqrt{p^2+M^2}}=\gamma M V(t),\] is the nonlinear drift term of the relativistic Orstein-Uhlenbeck model \cite{debbasch_relativistic_1997}, and $V(t)=p/\sqrt{p^2+M^2}<1$ is the relativistic velocity of the particle. Writing in terms of the absolute velocity, the interpretation of the drift term is simple: it is just the drift $-\gamma V$, but now with the velocity constrained by $V<c$.

For $M\gg p$ we find the classical case, while for $p\gg M$ we have the ultra-relativistic limit, where the velocity of the particle is close to the speed of light. The constants above are: the drift $\gamma$, the temperature of the surrounding $T$, and the rest mass of the particle $M$. The integrated noise $dW_t$ is a delta-correlated Wiener process \cite{wio2013path}, and models the fast degrees of freedom of the environment. According to \cite{koide_thermodynamic_2011,pal_stochastic_2020} the heat functional is given by
\begin{eqnarray}
    Q[p(t)]&=& \int_0^t \frac{p}{\sqrt{p^2+M^2}}\left(- \Gamma(p) +\sqrt{2\gamma MT}dW_\tau\right)d\tau \nonumber\\&=& \sqrt{p_t^2+M^2}-\sqrt{p_0^2+M^2}, 
\end{eqnarray}
where $Q<0$ means that the particle is losing energy to the environment, while $Q>0$ says that the particle is absorbing energy from the environment. The balance of energy is the same as for  the classical case. The drift term is responsible for the  negatives values, while the noise term yields positive values. What changes, comparing with the non-relativistic case, is the non-trivial dependence on $p$ in the drift term. Notice that  the first law of thermodynamics in the absence of work, i.e., $Q[p]=\Delta E$, is valid since $E=\sqrt{p^2+M^2}$ is the relativistic energy. 

We assume that the particle is initially in thermal equilibrium with the environment, with initial distribution given by the Juttner distribution \cite{cubero_thermal_2007}
\begin{equation}
    \rho(p_0,0) = \rho_0 \exp\left(-\beta \sqrt{p^2+M^2}\right),
\end{equation}
which is the correct equilibrium thermal distribution of the Relativistic Orstein-Uhlenbeck model, as verified by a microscopic collision simulation in \cite{cubero_thermal_2007}.

\subsection{Heat Functional}

The heat, being a functional of the trajectory, has the conditional probability
\begin{equation}
    P(Q|Q=Q[p])=\delta(Q-Q[p]).\label{cond}
\end{equation}
It emphasizes that the random values of $Q$ are given by the trajectory-dependent formula $Q[p]$. In the studied case, the heat only depends on the initial and final points of the trajectory. Therefore, the probability distribution for the heat will be given by
\begin{eqnarray}
    P(Q,t)&=&\int dp_t dp_0 \rho(p_0) \int_{p(0)=p_0}^{p(t)=p_t} Dp e^{-\mathcal{A}[p]}\delta(Q-Q[p(t)])\nonumber\\&=& \int dp_t P(Q,p_t,t),\label{pq}
\end{eqnarray}
where the path integral is over all the possible continuous non-differentiable trajectories \cite{chaichian2018path} and $\mathcal{A}[p(t)]$ is the stochastic action (see Appendix).

\section{Ultra-relativistic Regime}\label{sec3}

The ultra-relativistic regime occurs when the Brownian particle has a speed close to $c$, or equivalently $p\gg M$. In this regime the energy is given by a linear dependency in the momentum. As a consequence, the equation for the particle's momentum will be simplified. Interestingly, it has been shown that the ultra-relativistic Brownian motion, can describe the behavior of the charge carriers on a graphene plate \cite{pototsky_relativistic_2012,pototsky_periodically_2013} where, instead of the constant $c$, we have the Fermi velocity $v_f$. The analysis presented here can be cast as a simplified version of such a system as well.

\begin{figure}
    \includegraphics[width=8.6cm]{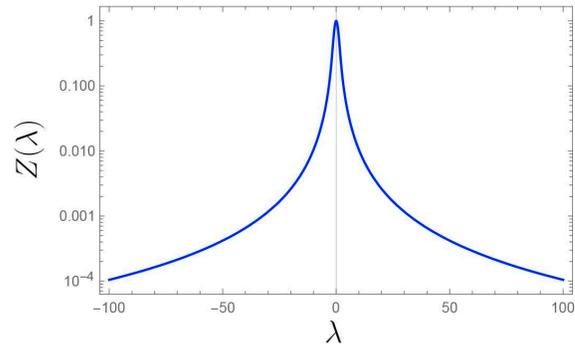}
    \caption{Characteristic function of the heat in $t=1$. Note that $Z(0)=1$. All constants are set to one.}
    \label{fig1}
\end{figure}

\begin{figure}
    \includegraphics[width=8.6cm]{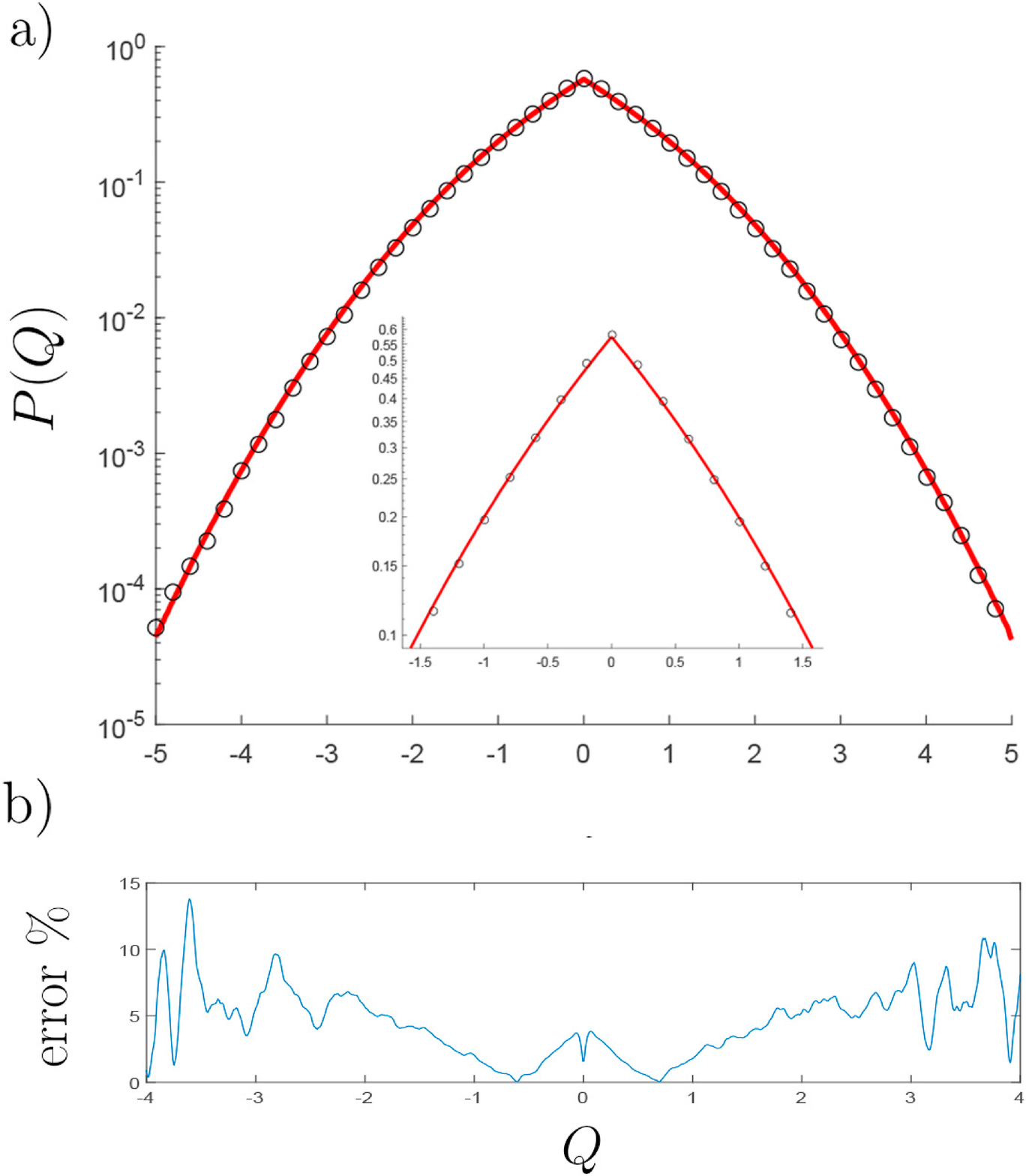}
    \caption{ a) Heat distribution for the ultra-relativistic particle at time $t=1$. The inset figure is the heat distribution in the small interval $[-1.5,1.5]$ . The solid red line is the theoretical solution, while the open circles are the numerical simulation b) Relative error in $P(Q)$ between simulation and solution. All constants are set to 1.}
    \label{fig2}
\end{figure}

In the ultra-relativistic regime, $E=c|p|$ and  the Langevin equation becomes
\begin{equation}
    dp = - \gamma c \;\frac{p}{|p|} dt + \sqrt{2\gamma T}dW_t,\label{langultra}
\end{equation}
 where the drift term is now similar to a dry friction force \cite{pototsky_periodically_2013,gennes_brownian_2005}. 
Notice that $c$ is no longer equals one here. (Because we don't have mass, it is important to use another parameter in the equations.) Following the definitions given in section \ref{sec2}, the heat exchanged between the particle and the environment will be
\begin{equation}
    Q[p]= \Delta E =c (|p_t|-|p_0|),
\end{equation}
which is the first law of Stochastic Thermodynamics \cite{koide_thermodynamic_2011,pal_stochastic_2020}. The heat distribution is then given by
\begin{equation}
    P(Q) = \int dp_t \int  dp_0 \rho(p_0) P[p_t,t|p_0]\delta(Q-Q[p]),
\end{equation}
where we remove the Dirac delta of the path integral in Eq.~\ref{pq}, making the path integral become the conditional probability. The conditional probability is written in as a path integral \cite{wio2013path,moreno_conditional_2019}
\begin{equation}
 P[p_t,t|p_0]= \int^{p(t)=p_t}_{p(0)=p_0} Dp\; e^{-\mathcal{A}[p(t)]},
\end{equation}
which is a sum over all trajectories that are continuous and non-differentiable \cite{chaichian2018path}. The result of this path integral is 
\begin{widetext}
\begin{equation}
    P[p_t,t|p_0]=\frac{e^{-\frac{c }{2 T}\left(-| p_0| +| p_t| +\frac{c \gamma  t}{2}\right)}}{\int P[p_t,t|p_0]dp_t} \left(\frac{c }{8 T}e^{\frac{c }{4 T}\left(\frac{c \gamma  t}{4}- (| p_0| +| p_t| )\right)} \left(\text{erf}\left(\frac{c \gamma  t-2 (| p_0| +| p_t| )}{4 \sqrt{\gamma  t T}}\right)+1\right)+\frac{e^{-\frac{(p_t-p_0)^2}{4 \gamma  t T}}}{2  \sqrt{\gamma  T\pi t}}\right),\label{ultratra}
\end{equation}
\end{widetext}
where $\text{erf}$ is the error function. We derive the above result in Appendix A. This conditional probability can be checked by numerical simulations of the Langevin Eq.~\ref{langultra}. 
We can rewrite the Dirac delta to find a more convenient formula for the heat distribution 
\begin{equation}
P(Q) = \int \frac{d\lambda}{2\pi} e^{i\lambda Q}Z(\lambda),\label{pqch}
\end{equation}
where
\begin{equation}
    Z(\lambda) = \int dp_t \int dp_0 \rho(p_0)  e^{-i\lambda c(|p_t|-|p_0|)}P[p_t,t|p_0],
\end{equation}
is the characteristic function of the heat, where we use the Juttner initial equilibrium distribution $\rho(p_0)=\rho_0\exp(-\beta c |p_0|)$. Due to the complicated dependence on $p_0$, $Z(\lambda)$ cannot be solved analytically. We find $Z(\lambda)$ by numerically integrating over $p_t$ and $p_0$. The result is plotted in Fig.\ref{fig1}. Notice that $Z(0)=1$, meaning that the distribution is properly normalized. Given $Z(\lambda)$ we use Eq.~\ref{pqch} to numerically integrate over $\lambda$ and find the heat distribution. The distribution is plotted in Fig.~\ref{fig2} a).

The heat distribution in Fig.~\ref{fig2} a) is symmetrical, meaning that the particle has no tendency to absorb or lose energy from the bath. This behavior is  encountered in its classical version, where $\Gamma(p)=-\gamma \frac{p}{M}$. It happens because we start in thermal equilibrium using the Juttner distribution. Moreover, compared with the classical case, the distribution is  smoother around $Q=0$, meaning that, for a  given trajectory, the chance for the particle absorbing or losing energy is larger than the classical case \cite{paraguassu_heat_2021_2}.

\section{Exact Solution}\label{sec4}
 To solve the model exactly,  a path integral technique can only give us an approximate result. Thus, we opt to use the Fokker-Planck formalism, which can be solved by numerical integration. We also compare the Fokker-Planck result with numerical simulations of the Langevin equation. The stochastic equation for the momentum is 
\begin{equation}
    dp = -\gamma M \frac{p}{\sqrt{p^2+M^2}}dt + \sqrt{2\gamma MT}dW_t. \label{rbm}
\end{equation}

\begin{figure}
    \includegraphics[width=8.6cm]{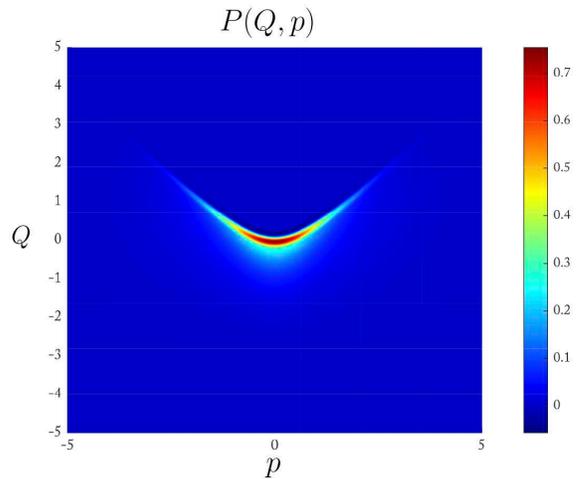}
    \caption{a) Joint distribution $P(Q,p)$. Note the shape of the curve obeys a behavior of $\sim\sqrt{(p^2+1)}$, which comes from the definition of heat. All constants are set to one.}
    \label{fig3}
\end{figure}

\begin{figure}
    \includegraphics[width=8.6cm]{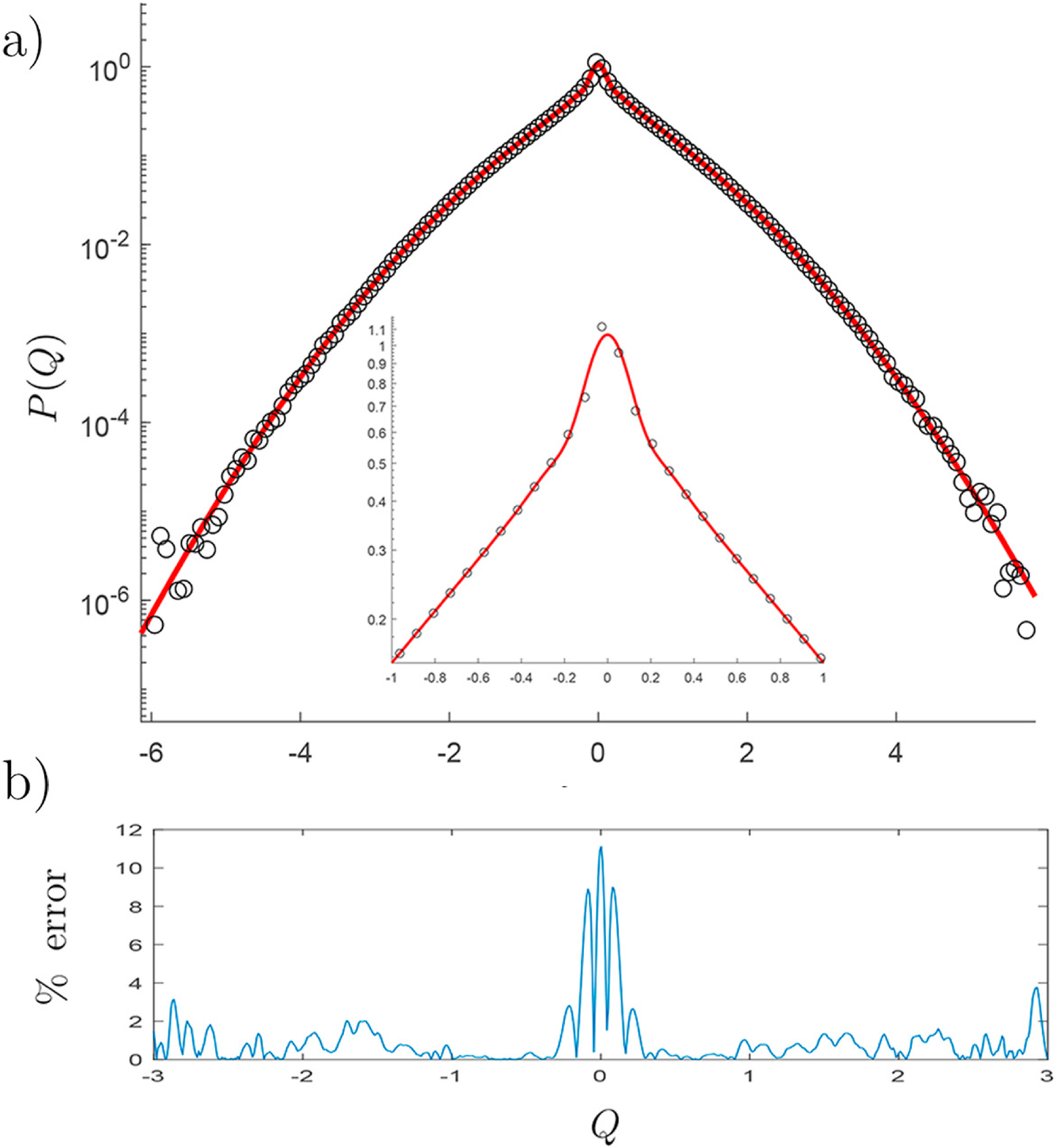}
    \caption{a)  Heat distribution in time $t=1$. Note the peak behavior in $Q=0$ a feature common in many stochastic thermodynamics systems. The inset figure is the heat distribution in the small interval $[-1,1]$. The red solid line is the Fokker-Planck result, while the open circles are the simulation of the Langevin equation. b) Relative error between simulation and numerical solution. All constants are set to one.}
    \label{fig4}
\end{figure}

The joint heat distribution $P(Q,p)$ can be solved exactly via the Fokker-Planck equation. One such approach was used in the derivation of the heat and work distributions of a Brownian particle in a double well potential \cite{imparato_work_2007,imparato_probability_2008}. One considers the stochastic equation for the heat
\begin{eqnarray}
    dQ= \frac{p}{\sqrt{p^2+M^2}} dp=\nonumber\\ =  -\gamma M\frac{p^2}{p^2+M^2} dt +\frac{p}{\sqrt{p^2+M^2}}\sqrt{2\gamma MT}dW_t,
\end{eqnarray}
where we replaced $dp$ by the Langevin equation Eq.~\ref{rbm}. With the above equation we can construct the Fokker-Planck for the joint distribution $P(Q,p)$. The Fokker-Planck for $P(Q,p)$ and its initial condition are 
\begin{eqnarray}
    \frac{\partial}{\partial t} P(Q,p) = -\nabla\cdot \left(\vec{F}P(Q,p)-\boldsymbol{B}\cdot\nabla P(Q,p)\right),\nonumber\\ P(Q,p,t=0) =\delta(Q) \rho(p_0),\label{fpjoint}
\end{eqnarray}
where $\nabla = \left(\partial_p,\partial_Q\right)$, and 
\begin{equation}
    \vec{F}=  \begin{pmatrix}
 \frac{-\gamma M p}{\sqrt{p^2+M^2}}\\
 \frac{-\gamma M p^2}{p^2+M^2} 
\end{pmatrix},\;\;\; \boldsymbol{B}=  \begin{pmatrix}
2\gamma M T    &  \frac{2\gamma M T p}{\sqrt{p^2+M^2}} \\
\frac{2\gamma M T p}{\sqrt{p^2+M^2}} & \frac{2\gamma M T p^2}{p^2+M^2}.  
\end{pmatrix} 
\end{equation}
Eq.~\ref{fpjoint} can be solved numerically giving $P(Q,p)$ which can be integrated numerically over $p$  to give the heat distribution $P(Q)$. We solve Eq.~\ref{fpjoint} numerically through the  Finite Elements method \cite{logg2012automated}, implemented by FEniCS \cite{langtangen2016solving}, which uses the variational formula of the PDE, also know as the weak form \cite{logg2012automated}. In our case, the variational formula is 
\begin{eqnarray}
    \int \partial_t P(Q,p)\; v d\Sigma = \nonumber\\=\int \vec{F}\cdot\nabla v P(Q,p) d\Sigma - \int \boldsymbol{B}\nabla P(Q,p) \cdot \nabla v d\Sigma,
\end{eqnarray}
where $v$ is a test function, and we are integrating in the domain $d\Sigma = dp dQ$. In this variational formula, the boundary condition in the weak form is  satisfied by 
\begin{equation}
    \oint (-\vec{F}P+\boldsymbol{B}\cdot\nabla P(Q,p))\cdot \hat{n} v\; d\partial\Sigma =0 
\end{equation}
which ensures the no flux boundary conditions over the integrated region.  We implement the time evolution by an Euler scheme. The result $P(Q,p)$ is plotted in Fig.~\ref{fig3}. With $P(Q,p)$ we integrate it numerically over $p$, obtaining the heat distribution $P(Q)$ plotted in Fig.~\ref{fig4} a). The relative error between the Fokker-Planck solution and the numerical simulation is in Fig.~\ref{fig4} c), showing the agreement between the two approaches.

The behavior of the heat distribution  is clear. The particle in average will not absorb or gain energy from the environment and it is equally probable to lose or to  gain energy since the distribution is symmetrical. This shows that the system is naturally close to equilibrium, which is expected since we assume an initially thermalized distribution. Moreover, comparing with the ultra-relativistic case, we see a peak around $Q=0$ meaning that the tendency of zero average heat is stronger than for the ultra-relativistic case.

\section{Conclusion}\label{sec5}
In the present paper, we have study the energy exchanged between a Relativistic diffusive particle and a thermal environment. We solve for the heat distribution in two different regimes, the ultra-relativistic and the exact relativistic limit. We first calculate the heat distribution for the ultra-relativistic limit through the use of path integrals. We find exact results through numerical integration of the characteristic function. For the exact limit, such an approach is not available, thus we use FEniCS \cite{langtangen2016solving} to numerically solve the Fokker-Planck of the joint distribution $P(Q,p)$, and then integrate over $p$ to find $P(Q)$. Both results are in essence exact and agree with numerical simulations. 

Physically, we find that in the Relativistic Stochastic Thermodynamics the heat shares similar statistical behavior compared with the classical (non-relativistic) case \cite{paraguassu_heat_2021_2} (see also \cite{rosinberg_stochastic_2017,munakata_entropy_2012}). The relativistic heat distribution for the free particle has a sharp distribution around $Q=0$, meaning that the particle neither absorbs nor gains energy on average from the environment. This result is also consistent with the classical case. Moreover, another similarity is the symmetry of the distribution, which is $P(Q)=P(-Q)$ showing that there is no preferred direction for the heat. One can expect this result since no external force or internal potential is interacting with the particle. Therefore, the proposed Relativistic Stochastic Thermodynamics leads to well-behaved physical properties for the heat.

The methods presented here can be generalized to higher dimensions, which is a more realistic scenario. A promising case is in the graphene chip \cite{pototsky_relativistic_2012}, wherre the charge carriers could obey the 2D Ultra-relativistic Langevin equation. Thus, one can study the protocols in the ultra-relativistic case, performing work on the system, and even study thermal machines protocols. Hence, the present study can  also serve as a starting point to the study of such machines. By calculating the heat, and work distribution, one can investigate the efficiency of such a case. This will be a theme of future work. 

Moreover, a more realistic scenario can also take into account interactions and multiplicative noise \cite{plyukhin_quasirelativistic_2013}. Both complications can also be treated with the methods exposed herein.

\section*{Acknowledgments}
We would like to thank Igor Brand\~ao, Juan Leite, and Victor Alencar for useful discussions. This work is supported by the Brazilian agencies CAPES and CNPq. P.V.P. would like to thank CNPq for his current fellowship. This study was financed in part by Coordena\c c\~ ao de Aperfei\c coamento de Pessoal de N\' ivel Superior - Brasil (CAPES) - Finance Code 001.

\appendix

\onecolumngrid
\section{Path integral for Ultra-relativistic case}

In equation \ref{ultratra} the conditional probability can be derived by means of path integral technique. Which states that
\begin{equation}
    P[p_t,t|p_0]= \int^{p(t)=p_t}_{p(0)=p_0} Dp\; e^{-\mathcal{A}[p(t)]}
\end{equation}
where $\mathcal{A}[p(t)]$ is the stochastic action \cite{wio2013path,chaichian2018path,moreno_conditional_2019}, in the Stratonovich prescription is given by
\begin{equation}
    \mathcal{A}[p(t)]= \frac{1}{4D}\int_0^t\left(\dot p+\gamma c \frac{p}{|p|}\right)^2d\tau -\frac{\gamma c}{2} \int_0^t \frac{\partial}{\partial p} \frac{p}{|p|} d\tau,
\end{equation}
where $D=\gamma T$.
By noticing that $p/|p|=\text{sign}(p)=2H(p)-1$ and $H'(p)=\delta(p)$, we can rewrite the action as
\begin{equation}
\mathcal{A}[p(t)]= \frac{1}{4D}\int_0^t\left(\dot{p}^2-\alpha \delta(p)\right)d\tau +\frac{\gamma c}{2D}\left(\frac{\gamma c }{2} t+|p_t|-|p_0|\right)
\end{equation}
where, $\alpha=4D\gamma c$. The conditional probability can be rewritten as
\begin{equation}
    P[p_t,t|p_0] = e^{-\frac{\gamma c}{2D}\left(\frac{\gamma c }{2} t+|p_t|-|p_0|\right)}K[p_t,t|p_0],
\end{equation}
where $K[p_t,t|p_0]$ will be the path integral
\begin{equation}
    K[p_t,t|p_0]= \int^{p(t)=p_t}_{p(0)=p_0} Dp \exp\left(-\frac{1}{4D}\int_0^t\left(\dot{p}^2-\alpha \delta(p)\right)d\tau\right)\label{prop}
\end{equation}
which has the same structure of a quantum mechanical propagator of a particle with a delta potential \cite{blinder_greens_1988,crandall_combinatorial_1993,goovaerts_new_1973,lawande_feynman_1988}. Thus, following \cite{lawande_feynman_1988} we review the derivation of this path integral. 

To solve Eq.~\ref{prop} we expand the potential, obtaining
\begin{equation}
    K[p_t,t|p_0] = K_0[p_t,t|p_0]+K_1[p_t,t|p_0]
\end{equation}
where
\begin{eqnarray}
K_0[p_t,t|p_0]= \int^{p(t)=p_t}_{p(0)=p_0} Dp e^{-\frac{1}{4D}\int \dot{p}^2d\tau}= \frac{1}{\sqrt{4\pi D t}}e^{-\frac{p_t^2}{4Dt}},\label{K0}\\
K_1[p_t,t|p_0]= \sum_{n=1}^{\infty} \left(\frac{-1}{4D}\right)^n\frac{(-\alpha)^n}{n!}\int Dp\; e^{-\frac{1}{4D}\int \dot{p}^2d\tau}\left(\int_0^t\delta(p)d\tau\right)^n.
\end{eqnarray}
Therefore, we only have to solve $K_1$. To do this, note that
\begin{eqnarray}
    \left(\int_0^t\delta(p)d\tau\right)^n = \int_0^t dt_n \int_0^t dt_{n-1} \dots \int_0^t dt_2 \int_0^t dt_1 \prod_{k=1}^n\delta(p(t_k))\\
    = n! \int_0^t dt_n \int_0^{t_n} dt_{n-1}\dots \int_0^{t_2} dt_1 \prod_{k=1}^n\delta(p(t_k))
\end{eqnarray}
where in the second line, we just reordered the time. The path integral in $K$ is a Wiener path integral that describes a Markovian stochastic process, thus we have the property
\begin{equation}
    K_0[p_t,t|p_0] = K_0[p_t,t|p_n,t_n] K_0[p_n,t_n|p_{n-1},t_{n-1}] \dots K_0[p_1,t_1|p_0,0],
\end{equation}
therefore, in $K_1$ we have 
\begin{eqnarray}
   \int dp_n \int dp_{n-1} \dots \int dp_1 K_0[p_t,t|p_n,t_n] \delta(p_n)\prod_{k=2}^{n} K_0[p_k,t_k|p_{k-1}t_{k-1}]K_0[p_1,t_1|p_0,0]\delta(p_k)\delta(p_1)=\\=K_0[p_t,t|0,t_n] \prod_{k=2}^{n} K_0[0,t_k|0,t_{k-1}]K_0[0,t_1|p_0,0].
\end{eqnarray}
Thus, we have
\begin{equation}
    K_1 = \sum_{n=1}^{\infty} \left(\frac{\alpha}{4D}\right)^n \int_0^t dt_n \int_0^{t_n} dt_{n-1}\dots \int_0^{t_2} dt_1 K_0[p_t,t|0,t_n] \prod_{k=2}^{n} K_0[0,t_k|0,t_{k-1}]K_0[0,t_1|p_0,0].
\end{equation}
Note that we have convolutions between the propagators. Then, by making the Laplace's transform
\begin{equation}
\tilde{K}_1 = \int_0^\infty e^{st} K_1[p_t,t|p_0],
\end{equation}
we can get rid of the time integrals, giving
\begin{equation}
    \tilde{K}_1  = \sum_{n=1}^{\infty} \left(\frac{\alpha}{4D}\right)^n \tilde{K}(p_t,s)\tilde{K}(0,s)^{n-1}\tilde{K}(p_0,s)
\end{equation}
where
\begin{eqnarray}
   \tilde{K}(p,s) = \int_0^\infty e^{st} \frac{e^{-\frac{p^2}{4Dt}}}{\sqrt{4\pi D t }} = \exp\left({-|p|\sqrt{\frac{s}{D}}}\right)\sqrt{\frac{1}{4\pi D s}},
\end{eqnarray}
then
\begin{equation}
    \tilde{K}_1  = \sum_{n=1}^{\infty} \left(\frac{\gamma c}{\sqrt{4 D}}\right)^n \exp\left({-(|p_t|+|p_0|)\sqrt{\frac{s}{D}}}\right)\left[\sqrt{\frac{1}{s}}\right]^{n+1}.
\end{equation}
The above sum is solved exactly, and then, we can use the inverse Laplace transform to find the desired $K_1$.
\begin{equation}
    K_1[p_t,t|p_0] = \frac{ c \gamma}{2 \pi i } \int_{\gamma -i\infty}^{\gamma+i\infty} \frac{\exp\left({-(|p_t|+|p_0|)\sqrt{\frac{s}{D}}}+st\right)}{\sqrt{sD}(2\sqrt{sD}-\gamma c)} ds
\end{equation}
we can make $s=w^2$, giving
\begin{equation}
  K_1[p_t,t|p_0]  = \frac{ c \gamma}{ i \pi \sqrt{D}} \int_C \frac{\exp\left({-(|p_t|+|p_0|)\sqrt{\frac{1}{D}}w}+w^2t\right)}{(2\sqrt{D}w-\gamma c)}dw
\end{equation}
rewriting the denominator as an integral
\begin{equation}
 K_1[p_t,t|p_0]  = \frac{ c \gamma}{i\pi\sqrt{D}  } \int_0^\infty\int_C {\exp\left({-(|p_t|+|p_0|)\sqrt{\frac{1}{D}}w}+w^2t+z(2\sqrt{D}w-\gamma c)\right)}dwdz
\end{equation}
This allow us to solve the integral in $w$ as a gaussian integral, giving
\begin{equation}
    K_1[p_t,t|p_0]  = \frac{ c \gamma} {\sqrt{\pi t}}\int_0^\infty e^{\frac{z \gamma c}{4D}} e^{-\frac{(z+\xi)^2}{4Dt}} dz,
\end{equation}
where $\xi = |p_t|+|p_0|$. We finally have
\begin{equation}
    K_1[p_t,t|p_0] =\frac{\gamma c }{8 D}e^{\frac{\gamma c }{4 D}\left(\frac{c \gamma  t}{4}- \xi \right)} \left(\text{erf}\left(\frac{c \gamma  t-2 \xi }{4 \sqrt{D t }}\right)+1\right).\label{K1}
\end{equation}

Finally, with Eq.~\ref{K1} and \ref{K0} we have the conditional probability
\begin{equation}
    P[p_t,t|p_0] = e^{-\frac{c }{2 T}\left(-| p_0| +| p_t| +\frac{c \gamma  t}{2}\right)}\left(\frac{\gamma c }{8 D}e^{\frac{c }{4 T}\left(\frac{c \gamma  t}{4}- \xi\right)} \left(\text{erf}\left(\frac{c \gamma  t-2 \xi}{4 \sqrt{Dt }}\right)+1\right)+\frac{e^{-\frac{(p_t-p_0)^2}{4 D t }}}{2  \sqrt{D\pi t}}\right),
\end{equation}
which can be normalized, giving Eq.~\ref{cond}.

\section*{References}

\twocolumngrid

\bibliography{name.bib}

\providecommand{\newblock}{}
\begin{thebibliography}{10}
\expandafter\ifx\csname url\endcsname\relax
  \def\url#1{{\tt #1}}\fi
\expandafter\ifx\csname urlprefix\endcsname\relax\def\urlprefix{URL }\fi
\providecommand{\eprint}[2][]{\url{#2}}

\bibitem{carnot2012reflections}
Carnot S 2012 {\em Reflections on the motive power of fire: And other papers on
  the second law of thermodynamics\/} (Courier Corporation)

\bibitem{oliveira2020classical}
Oliveira M~J~d 2020 {\em Revista Brasileira de Ensino de F{\'\i}sica\/} {\bf
  42}

\bibitem{ciliberto_experiments_2017}
Ciliberto S 2017 {\em Phys. Rev. X\/} {\bf 7} 021051 publisher: American
  Physical Society
  \urlprefix\url{https://link.aps.org/doi/10.1103/PhysRevX.7.021051}

\bibitem{seifert2012stochastic}
Seifert U 2012 {\em Reports on progress in physics\/} {\bf 75} 126001

\bibitem{sekimoto2010stochastic}
Sekimoto K 2010 {\em Stochastic energetics\/} vol 799 (Springer)

\bibitem{paraguassu_heat_2021_2}
Paraguass{\'u} P~V, Aquino R and Morgado W~A~M 2021 {\em arXiv:2102.09115
  [cond-mat]\/} ArXiv: 2102.09115
  \urlprefix\url{http://arxiv.org/abs/2102.09115}

\bibitem{paraguassu_heat_2021}
Paraguass{\'u} P~V and Morgado W~A~M 2021 {\em J. Stat. Mech.\/} {\bf 2021}
  023205 ISSN 1742-5468 publisher: IOP Publishing
  \urlprefix\url{https://doi.org/10.1088/1742-5468/abda25}

\bibitem{gupta_heat_2021}
Gupta D and Sivak D~A 2021 {\em arXiv:2103.09358 [cond-mat]\/} ArXiv:
  2103.09358 \urlprefix\url{http://arxiv.org/abs/2103.09358}

\bibitem{fogedby_heat_2020}
Fogedby H~C 2020 {\em J. Stat. Mech.\/} {\bf 2020} 083208 ISSN 1742-5468
  publisher: IOP Publishing
  \urlprefix\url{https://doi.org/10.1088/1742-5468/aba7b2}

\bibitem{goswami_heat_2019}
Goswami K 2019 {\em Phys. Rev. E\/} {\bf 99} 012112 publisher: American
  Physical Society
  \urlprefix\url{https://link.aps.org/doi/10.1103/PhysRevE.99.012112}

\bibitem{crisanti_heat_2017}
Crisanti A, Sarracino A and Zannetti M 2017 {\em Phys. Rev. E\/} {\bf 95}
  052138 publisher: American Physical Society
  \urlprefix\url{https://link.aps.org/doi/10.1103/PhysRevE.95.052138}

\bibitem{ghosal_distribution_2016}
Ghosal A and Cherayil B~J 2016 {\em J. Stat. Mech.\/} {\bf 2016} 043201 ISSN
  1742-5468 publisher: IOP Publishing
  \urlprefix\url{https://doi.org/10.1088/1742-5468/2016/04/043201}

\bibitem{rosinberg_heat_2016}
Rosinberg M~L, Tarjus G and Munakata T 2016 {\em EPL\/} {\bf 113} 10007 ISSN
  0295-5075 publisher: IOP Publishing
  \urlprefix\url{https://doi.org/10.1209/0295-5075/113/10007}

\bibitem{kim_heat_2014}
Kim K, Kwon C and Park H 2014 {\em Phys. Rev. E\/} {\bf 90} 032117 ISSN
  1539-3755, 1550-2376
  \urlprefix\url{https://link.aps.org/doi/10.1103/PhysRevE.90.032117}

\bibitem{kusmierz_heat_2014}
Ku{\'s}mierz {\textbackslash}, Rubi J~M and Gudowska-Nowak E 2014 {\em J. Stat.
  Mech.\/} {\bf 2014} P09002 ISSN 1742-5468 publisher: IOP Publishing
  \urlprefix\url{https://doi.org/10.1088/1742-5468/2014/09/p09002}

\bibitem{saha_work_2014}
Saha B and Mukherji S 2014 {\em J. Stat. Mech.\/} {\bf 2014} P08014 ISSN
  1742-5468
  \urlprefix\url{https://iopscience.iop.org/article/10.1088/1742-5468/2014/08/P08014}

\bibitem{chatterjee_single-molecule_2011}
Chatterjee D and Cherayil B~J 2011 {\em J. Stat. Mech.\/} {\bf 2011} P03010
  ISSN 1742-5468
  \urlprefix\url{https://iopscience.iop.org/article/10.1088/1742-5468/2011/03/P03010}

\bibitem{chatterjee_exact_2010}
Chatterjee D and Cherayil B~J 2010 {\em Phys. Rev. E\/} {\bf 82} 051104 ISSN
  1539-3755, 1550-2376
  \urlprefix\url{https://link.aps.org/doi/10.1103/PhysRevE.82.051104}

\bibitem{fogedby_heat_2009}
Fogedby H~C and Imparato A 2009 {\em J. Phys. A: Math. Theor.\/} {\bf 42}
  475004 ISSN 1751-8121 publisher: IOP Publishing
  \urlprefix\url{https://doi.org/10.1088/1751-8113/42/47/475004}

\bibitem{imparato_probability_2008}
Imparato A, Jop P, Petrosyan A and Ciliberto S 2008 {\em J. Stat. Mech.\/} {\bf
  2008} P10017 ISSN 1742-5468
  \urlprefix\url{https://iopscience.iop.org/article/10.1088/1742-5468/2008/10/P10017}

\bibitem{imparato_work_2007}
Imparato A, Peliti L, Pesce G, Rusciano G and Sasso A 2007 {\em Phys. Rev. E\/}
  {\bf 76} 050101 publisher: American Physical Society
  \urlprefix\url{https://link.aps.org/doi/10.1103/PhysRevE.76.050101}

\bibitem{joubaud_fluctuation_2007}
Joubaud S, Garnier N~B and Ciliberto S 2007 {\em J. Stat. Mech.\/} {\bf 2007}
  P09018--P09018 ISSN 1742-5468 publisher: IOP Publishing
  \urlprefix\url{https://doi.org/10.1088/1742-5468/2007/09/p09018}

\bibitem{pal_stochastic_2020}
Pal P~S and Deffner S 2020 {\em New J. Phys.\/} {\bf 22} 073054 ISSN 1367-2630
  \urlprefix\url{https://iopscience.iop.org/article/10.1088/1367-2630/ab9ce6}

\bibitem{meyer2018cosmic}
Meyer E~T 2018 {\em Nature Astronomy\/} {\bf 2} 32--33

\bibitem{koide_thermodynamic_2011}
Koide T and Kodama T 2011 {\em Phys. Rev. E\/} {\bf 83} 061111 ISSN 1539-3755,
  1550-2376 \urlprefix\url{https://link.aps.org/doi/10.1103/PhysRevE.83.061111}

\bibitem{akamatsu_heavy_2009}
Akamatsu Y, Hatsuda T and Hirano T 2009 {\em Phys. Rev. C\/} {\bf 79} 054907
  ISSN 0556-2813, 1089-490X
  \urlprefix\url{https://link.aps.org/doi/10.1103/PhysRevC.79.054907}

\bibitem{pototsky_relativistic_2012}
Pototsky A, Marchesoni F, Kusmartsev F~V, H{\"a}nggi P and
  Savel{\textquoteright}ev S~E 2012 {\em Eur. Phys. J. B\/} {\bf 85} 356 ISSN
  1434-6028, 1434-6036
  \urlprefix\url{http://link.springer.com/10.1140/epjb/e2012-30716-7}

\bibitem{pototsky_periodically_2013}
Pototsky A and Marchesoni F 2013 {\em Phys. Rev. E\/} {\bf 87} 032132 ISSN
  1539-3755, 1550-2376
  \urlprefix\url{https://link.aps.org/doi/10.1103/PhysRevE.87.032132}

\bibitem{dunkel_theory_2005}
Dunkel J and H{\"a}nggi P 2005 {\em Phys. Rev. E\/} {\bf 71} 016124 ISSN
  1539-3755, 1550-2376
  \urlprefix\url{https://link.aps.org/doi/10.1103/PhysRevE.71.016124}

\bibitem{dunkel_theory_2005-1}
Dunkel J and H{\"a}nggi P 2005 {\em Phys. Rev. E\/} {\bf 72} 036106 ISSN
  1539-3755, 1550-2376
  \urlprefix\url{https://link.aps.org/doi/10.1103/PhysRevE.72.036106}

\bibitem{dunkel_relativistic_2009}
Dunkel J and H{\"a}nggi P 2009 {\em Physics Reports\/} {\bf 471} 1--73 ISSN
  03701573
  \urlprefix\url{https://linkinghub.elsevier.com/retrieve/pii/S0370157308004171}

\bibitem{lindner_diffusion_2007}
Lindner B 2007 {\em New J. Phys.\/} {\bf 9} 136--136 ISSN 1367-2630
  \urlprefix\url{https://iopscience.iop.org/article/10.1088/1367-2630/9/5/136}

\bibitem{debbasch_relativistic_1997}
Debbasch F, Mallick K and Rivet J~P 1997 {\em Journal of Statistical Physics\/}
  {\bf 88} 945--966 ISSN 1572-9613
  \urlprefix\url{https://doi.org/10.1023/B:JOSS.0000015180.16261.53}

\bibitem{debbasch_thermal_2012}
Debbasch F, Espaze D, Foulonneau V and Rivet J~P 2012 {\em Physica A:
  Statistical Mechanics and its Applications\/} {\bf 391} 3797--3804 ISSN
  03784371
  \urlprefix\url{https://linkinghub.elsevier.com/retrieve/pii/S0378437112001732}

\bibitem{felderhof_momentum_2012}
Felderhof B~U 2012 {\em Phys. Rev. E\/} {\bf 86} 061103 ISSN 1539-3755,
  1550-2376 \urlprefix\url{https://link.aps.org/doi/10.1103/PhysRevE.86.061103}

\bibitem{langtangen2016solving}
Langtangen H~P and Logg A 2016 {\em Solving PDEs in python: the FEniCS tutorial
  I\/} (Springer Nature)

\bibitem{risken1996fokker}
Risken H 1996 Fokker-planck equation {\em The Fokker-Planck Equation\/}
  (Springer) pp 63--95

\bibitem{onsager1953fluctuations}
Onsager L and Machlup S 1953 {\em Physical Review\/} {\bf 91} 1505

\bibitem{moreno_conditional_2019}
Moreno M~V, Barci D~G and Arenas Z~G 2019 {\em Phys. Rev. E\/} {\bf 99} 032125
  ISSN 2470-0045, 2470-0053
  \urlprefix\url{https://link.aps.org/doi/10.1103/PhysRevE.99.032125}

\bibitem{wio2013path}
Wio H~S 2013 {\em Path integrals for stochastic processes: An introduction\/}
  (World Scientific)

\bibitem{chaichian2018path}
Chaichian M and Demichev A 2018 {\em Path integrals in physics: Volume I
  stochastic processes and quantum mechanics\/} (CRC Press)

\bibitem{cubero_thermal_2007}
Cubero D, Casado-Pascual J, Dunkel J, Talkner P and H{\"a}nggi P 2007 {\em
  Phys. Rev. Lett.\/} {\bf 99} 170601 ISSN 0031-9007, 1079-7114
  \urlprefix\url{https://link.aps.org/doi/10.1103/PhysRevLett.99.170601}

\bibitem{gennes_brownian_2005}
Gennes P~G~d 2005 {\em J Stat Phys\/} {\bf 119} 953--962 ISSN 0022-4715,
  1572-9613 \urlprefix\url{http://link.springer.com/10.1007/s10955-005-4650-4}

\bibitem{logg2012automated}
Logg A, Mardal K~A and Wells G 2012 {\em Automated solution of differential
  equations by the finite element method: The FEniCS book\/} vol~84 (Springer
  Science \& Business Media)

\bibitem{rosinberg_stochastic_2017}
Rosinberg M~L, Tarjus G and Munakata T 2017 {\em Phys. Rev. E\/} {\bf 95}
  022123 ISSN 2470-0045, 2470-0053
  \urlprefix\url{https://link.aps.org/doi/10.1103/PhysRevE.95.022123}

\bibitem{munakata_entropy_2012}
Munakata T and Rosinberg M~L 2012 {\em J. Stat. Mech.\/} {\bf 2012} P05010 ISSN
  1742-5468 publisher: IOP Publishing
  \urlprefix\url{https://iopscience.iop.org/article/10.1088/1742-5468/2012/05/P05010/meta}

\bibitem{plyukhin_quasirelativistic_2013}
Plyukhin A~V 2013 {\em Phys. Rev. E\/} {\bf 88} 052115 ISSN 1539-3755,
  1550-2376 \urlprefix\url{https://link.aps.org/doi/10.1103/PhysRevE.88.052115}

\bibitem{blinder_greens_1988}
Blinder S~M 1988 {\em Phys. Rev. A\/} {\bf 37} 973--976 publisher: American
  Physical Society
  \urlprefix\url{https://link.aps.org/doi/10.1103/PhysRevA.37.973}

\bibitem{crandall_combinatorial_1993}
Crandall R~E 1993 {\em J. Phys. A: Math. Gen.\/} {\bf 26} 3627--3648 ISSN
  0305-4470 publisher: IOP Publishing
  \urlprefix\url{https://doi.org/10.1088/0305-4470/26/14/024}

\bibitem{goovaerts_new_1973}
Goovaerts M~J, Babcenco A and Devreese J~T 1973 {\em Journal of Mathematical
  Physics\/} {\bf 14} 554--559 ISSN 0022-2488 publisher: American Institute of
  Physics \urlprefix\url{https://aip.scitation.org/doi/10.1063/1.1666355}

\bibitem{lawande_feynman_1988}
Lawande S~V and Bhagwat K~V 1988 {\em Physics Letters A\/} {\bf 131} 8--10 ISSN
  0375-9601
  \urlprefix\url{https://www.sciencedirect.com/science/article/pii/0375960188906226}

\end{thebibliography}
\bibliographystyle{name.bst}

\end{document}